\documentclass[sigconf]{acmart}

\renewcommand\footnotetextcopyrightpermission[1]{} 

 \setcopyright{none}

\usepackage{color}
\usepackage{url}
\usepackage{tikz}
\usepackage{verbatim} 
\usepackage{subfigure} 

\usepackage{multirow}
\usepackage{algorithm}
\usepackage[noend]{algorithmic}
\usepackage{xspace}
\usepackage{graphicx}
\usepackage{epsfig}
\usepackage{amsmath}
\usepackage{amssymb}
\usepackage{float}
\usepackage{enumitem}
\usepackage{balance}
\usepackage[font=bf,belowskip=0pt,aboveskip=5pt]{caption}
\usepackage{lipsum}

\usepackage{amsmath}
\makeatletter
\g@addto@macro\normalsize{%
  \setlength\abovedisplayskip{2pt}
  \setlength\belowdisplayskip{2pt}
  \setlength\abovedisplayshortskip{2pt}
  \setlength\belowdisplayshortskip{2pt}
}

\usepackage{epigraph} 

\usepackage{etoolbox}
\apptocmd{\thebibliography}{\normalsize}{}{}

\usepackage{booktabs} 

\setitemize{noitemsep,topsep=0pt,parsep=0pt,partopsep=0pt,leftmargin=*}

\newcommand{\hide}[1]{}
\newcommand{\xhdr}[1]{\vspace{1.7mm}\noindent{{\bf #1.}}}

\newcommand{\eg}{\emph{e.g.}}
\newcommand{\ie}{\emph{i.e.}}

\graphicspath{{./FIG/}}

\setitemize{noitemsep,topsep=0pt,parsep=0pt,partopsep=0pt,leftmargin=*}

\begin{document}

\copyrightyear{2018}
\acmYear{2018}
\setcopyright{iw3c2w3}
\acmConference[WWW 2018]{The 2018 Web Conference}{April 23--27, 2018}{Lyon,
France}
\acmBooktitle{WWW 2018: The 2018 Web Conference, April 23--27, 2018, Lyon, France}
\acmPrice{}
\acmDOI{10.1145/3178876.3186052}
\acmISBN{978-1-4503-5639-8}

\title{Modeling Individual Cyclic Variation in Human Behavior}
\author{Emma Pierson}
\affiliation{%
  \institution{Stanford University}
}
\email{emmap1@cs.stanford.edu}

\author{Tim Althoff}
\affiliation{%
  \institution{Stanford University}
}
\email{althoff@cs.stanford.edu}

\author{Jure Leskovec}
\affiliation{%
  \institution{Stanford University}
}
\email{jure@cs.stanford.edu}

\renewcommand{\shorttitle}{Modeling Individual Cyclic Variation in Human Behavior}

\begin{abstract}

Cycles are fundamental to human health and behavior. Examples include mood cycles, circadian rhythms, and the menstrual cycle. However, modeling cycles in time series data is challenging because in most cases the cycles are not labeled or directly observed and need to be inferred from multidimensional measurements taken over time. Here, we present \emph{CyHMMs}, a cyclic hidden Markov model method for detecting and modeling cycles in a collection of multidimensional heterogeneous time series data. In contrast to previous cycle modeling methods, CyHMMs deal with a number of challenges encountered in modeling real-world cycles: they can model multivariate data with both discrete and continuous dimensions; they explicitly model and are robust to missing data; and they can share information across individuals to accommodate variation both within and between individual time series.

Experiments on synthetic and real-world health-tracking data demonstrate that CyHMMs infer cycle lengths more accurately than existing methods, with 58\% lower error on simulated data and 63\% lower error on real-world data compared to the best-performing baseline. CyHMMs can also perform functions which baselines cannot: they can model the progression of individual features/symptoms over the course of the cycle, identify the most variable features, and cluster individual time series into groups with distinct characteristics. Applying CyHMMs to two real-world health-tracking datasets---of human menstrual cycle symptoms and physical activity tracking data---yields important insights including which symptoms to expect at each point during the cycle. We also find that people fall into several groups with distinct cycle patterns, and that these groups differ along dimensions not provided to the model. For example, by modeling missing data in the menstrual cycles dataset, we are able to discover a medically relevant group of birth control users even though information on birth control is not given to the model. 

\end{abstract}

\maketitle

\section{Introduction}
\label{sec:intro}
Detecting and modeling cyclic patterns is important for understanding human health and behavior \cite{han1999efficient}. Cycles appear in domains as varied as psychology (\eg, mood cycles \cite{golder2011diurnal} and cyclic mood disorders \cite{findling2001rapid, partonen1998seasonal}); physiology (\eg, circadian rhythms and hormonal cycles \cite{althoff2017harnessing, chiazze1968length}); and biology (\eg, the cell cycle \cite{mitchison1971biology}). Modeling such cycles enables beneficial health interventions: modeling mood cycles can predict symptoms and guide treatments \cite{mc2016seasonal}, and modeling circadian cycles can predict fatigue and improve safety \cite{dinges1995overview,althoff2017harnessing}. 

In the cycles described above, a latent state progresses cyclically and influences observed data. Consider for concreteness the running example of hormone cycles. A person's hormonal state---that is, the levels of hormones in their bloodstream---is generally unobserved, but it progresses cyclically and can influence observed behavior~\cite{pearlstein2005pretreatment}. For example, in women, a single hormone cycle consists of moving through a series of states---follicular, late follicular, midluteal, late luteal \cite{berga1990circadian}---before returning to the follicular state. (Throughout this paper, completing a \emph{cycle} means progressing through all latent states in a defined order before returning to the original latent state.) Modeling hormone cycles includes several problems of interest in human health: How does one compute the length of a cycle for a particular person? Can we cluster individuals into groups with distinct cycle dynamics, who may benefit from different medical interventions? How do features/symptoms vary throughout the cycle, since features that increase at the same point in the cycle may have the same physiological cause?  (We use \emph{feature} to refer to a single observed dimension of a multivariate time series: in the hormone cycles case, a binary feature might be whether the person reported negative mood on each day, and a second feature might be whether they reported pain).

Modeling cycles presents a number of challenges. The cycle state for each individual at each timestep is generally unobserved: a person's cyclic hormonal state may influence the level of positive affect they express on social media, but social media data is rarely linked to data on hormone state. In addition, cyclic dynamics vary across individuals---for example, individuals may experience cycles of slightly different lengths---and vary over time within a single individual as well. Further, real-world time series data is multivariate with discrete as well as continuous features and missing data. 

Numerous methods have been developed to model cyclic patterns, including Fourier transforms, autocorrelation, data mining techniques, and computational biology algorithms ~\cite{bracewell1986fourier, chaovalit2011discrete, daubechies1990wavelet, pindyck1998econometric, aref2004incremental, han1999efficient, ma2001mining, yang2010analyzing}. However, these techniques fail to address one or more of the challenges described above and, because they are not generative models, lack the ability to fully model the cycle: for example, they are not designed to model how each feature varies throughout the cycle or cluster individuals into distinct groups. In addition, previous work has been limited by the lack of ground-truth data on the true latent cycle state, which has prevented quantitative evaluation and comparisons between methods. That is, if an unsupervised method claims that cycles have a certain length or a certain feature or symptom varies particularly dramatically over the course of the cycle, how do we assess whether  those inferences are accurate? 

\xhdr{This work}
Here we present \emph{CyHMMs}, a cyclic hidden Markov model method which models cycles and addresses the challenges described above. CyHMMs take as input a collection of multivariate time series, with one time series for every individual in a population; time series can have missing data and discrete or continuous values. For each individual and each timestep, CyHMMs infer a discrete latent cycle state from the observed time series, using the fact that features vary dynamically based on the latent cycle state. Given these inferred cycle states, CyHMMs can then (1) compute the \emph{cycle length} as the time it takes an individual to return to the same latent state, (2) predict \emph{feature trajectories} by using the fitted generative model to predict how each feature will change over the course of the cycle, (3) identify the \emph{most variable features}, and (4) \emph{cluster} individuals into groups with distinct cycle dynamics.

We evaluate CyHMMs on both simulated data and two real-world cyclic datasets---of human menstrual cycle data, and activity tracking data---where we have knowledge of the true cycle state.  Using simulated data with known parameters, we show that CyHMMs recover the cycle length 58\% more accurately than the best performing baseline.  
On real-world data, CyHMMs recover the cycle length 63\% more accurately than the best performing baseline. Furthermore, we show that CyHMMs can accurately infer other fundamental aspects of the cycle which existing methods cannot: 
\begin{itemize}
\item \emph{Feature trajectories:} CyHMMs can infer how each feature or symptom fluctuates over the course of the cycle. This is important for understanding an individual's likely time course through the cycle, and for understanding which features tend to peak during the same cycle state, potentially indicating a common cause. 
\item \emph{Feature variability:} CyHMMs can infer which features are most variable. This is important for understanding which features are affected by the cycle, and which remain relatively constant. 
\item \emph{Clustering:} CyHMMs can partition individuals into groups which correlate with true population heterogeneity. This is important because not all individuals have the same cycle dynamics, and so different individuals may benefit from different interventions. 
\end{itemize}

Our inferences on two real-world datasets further demonstrate the utility of cycle modeling in the important human health domains of activity tracking, where we show we can accurately recover weekly sleep cycles, and menstrual cycle tracking, where we show we can accurately recover the 28-day cycle. 
For example, in the menstrual cycle data, CyHMMs identify a subpopulation of users who are much more likely to be taking hormonal birth control---even though no information on birth control is provided to the model. Identifying such subpopulations is essential for accurately characterizing cycles (since women taking birth control experience different menstrual cycle dynamics \cite{oinonen2002extent, paige1971effects}) and to designing medical treatments (since women taking birth control require different pharmacologic interventions \cite{hassan1987pharmacologic} and have different risks of hormone-related cancers \cite{collaborative1996breast,narod1998oral}). We emphasize that while in this paper we analyze datasets where cycle states are known in order to prove that CyHMMs work, CyHMMs are generally applicable to time series where cyclic dynamics are suspected but cycle states are not known. Our CyHMM implementation is publicly available.

\section{Related Work}
\label{sec:related}

\label{previous_work}
Our work draws on an extensive literature of methods developed for detecting and modeling cyclic patterns; we briefly describe major approaches. \label{label:related_work} \emph{Fourier transforms} \cite{bracewell1986fourier} express a time-dependent signal as a sum of sinusoids or complex exponentials; the most significant frequencies in the signal can be extracted from the amplitudes of the Fourier coefficients. While Fourier transforms are a classic technique for univariate time series, they are not easily adapted to data which is multivariate, missing, or discrete. Further, because they are not generative models, they cannot easily be used to predict how features will vary over the course of the cycle. It is also not clear how to apply them to a population of individuals whose time series are different but related. Fourier transforms also assume that the cycle length in each signal remains constant, which is not true for individuals whose cycle lengths vary from cycle to cycle. (Wavelet transforms \cite{daubechies1990wavelet} overcome this last deficiency but not the other ones.)  Another common technique for detecting periodicity, \emph{autocorrelation}, computes the correlation between a time series and a lagged copy of itself \cite{pindyck1998econometric}. It is mathematically related to the Fourier transform and suffers from similar drawbacks.

More recent \emph{data mining techniques} find ``partial periodic patterns'' that occur repeatedly in time series, sometimes with imperfect regularity \cite{aref2004incremental, han1999efficient, han1998mining, ma2001mining, berberidis2002discovery, kiran2015discovering, chanda2015efficient, hu2015novel, li2015eperiodicity,han2004mining,giannella2003mining,yang2003mining,yang2001infominer}. These methods represent discrete time series as strings and search for recurring patterns. Han et al. \cite{han1998mining, han1999efficient} introduce the concept of partial periodicity; numerous related approaches have been developed, including Ma and Hellerstein's algorithm for detecting periodic events with unknown periods~\cite{ma2001mining} and methods for detecting frequently recurring patterns~\cite{han2004mining, giannella2003mining, orlando2002adaptive}. These models do not address several challenges addressed in this work. They are not full generative models and do not recover all the cycle parameters in which we are interested; they cannot, for example, predict how observed features will vary over the course of the cycle. Second, we seek a method applicable to \emph{both} discrete and continuous data, and string-based methods are inherently discrete. 

There is also a relevant literature in computational biology which searches for circadian rhythms, often in genomic data \cite{yang2010analyzing, deckard2013design, wu2014evaluation, hughes2010jtk_cycle}, by fitting a periodic model to individual genes. These models have several drawbacks for the real-world datasets we consider: they are designed for data which is much higher-dimensional than most human activity datasets; they are designed for continuous data, and may not work well on discrete data; they consider each feature individually rather than sharing information across features; and they do not easily allow sharing of information across individuals. 

Because our method infers a hidden state at each timestep, it is also inspired by previous approaches which use latent states---and hidden Markov models specifically---to model time series. Such latent state approaches have been applied to model time series in domains such as speech recognition \cite{rabiner1989tutorial}, healthcare \cite{wang2014unsupervised} and sports \cite{kostakis2017discovering}. These approaches have been successful because multivariate time series in many domains are often generated by low-dimensional hidden states, motivating our application to cyclic time series. 
HMM models that contain cyclic topologies have diverse applications: for example, in video analysis \cite{magee2002detecting}, speaker recognition \cite{zheng1988text}, shape descriptors \cite{arica2000shape}, arrythmia detection \cite{coast1990approach}, and protein sequence analysis \cite{sonnhammer1998hidden}.
 Finally, our work draws on ideas from hidden semi-Markov models (HSMMs) and explicit duration hidden Markov models (EDHMMs) \cite{levinson1986continuously, ferguson1980variable, yu2010hidden} which develop formalisms for modeling latent states with non-geometric duration distributions that are commonly found in real-world data~\cite{althoff2017harnessing,lenton1984normal}.

\section{Task Description}
\label{sec:task}

We assume the data consist of $N$ multivariate time series, one for each individual $i$ in a population. Throughout, we use \emph{feature} to refer to a single dimension of a multivariate time series. Each time series, $X^{(i)}$, is a $T^{(i)} \times K$ matrix, where $T^{(i)}$ is the number of timesteps in the time series and $K$ is the number of features. $X^{(i)}_{tk}$ denotes the value in the $i$-th time series at the $t$-th timestep  of the $k$-th feature. Features may be binary or continuous and may have missing data. Our task is to infer basic cycle characteristics:

\begin{itemize}
\item \emph{Cycle length:} both for each individual and the whole population. 
\item \emph{Feature progression:} how each feature varies over the course of the cycle. 
\item \emph{Feature variability:} which features vary the most over the course of the cycle and which remain relatively constant. 
\item \emph{Clustering:} whether the population can be divided into groups with distinct cycle characteristics. 
\end{itemize}

\section{Proposed Model}
\label{sec:model}

\label{model_description}

We propose \emph{CyHMMs}, a cyclic hidden Markov model method for modeling cyclical time series data. As in a standard HMM, at each timestep an individual is in one of $J$ latent states $Z_j$, indexed by $j$, and emits an observed feature vector drawn from a distribution specific to that latent state. However, CyHMMs differ from standard HMMs in two ways. First, they capture cyclicity by placing constraints on how individuals can transition between states: each individual must progress through states $Z_1, Z_2, ... $ in the same order (although they can begin in any start state) and after reaching the final state $Z_J$ must return to $Z_1$ to begin the cycle again. An individual thus completes a \emph{cycle} when they pass through all latent states and return to the same latent state. We constrain latent state progression in this way because it naturally encodes how the latent state in real-world cycles progresses cyclically and because it makes the model interpretable. If the latent state transitions were unconstrained it would not be clear how to define cycles.

\begin{figure}[t!]
\includegraphics[width=.8\columnwidth]{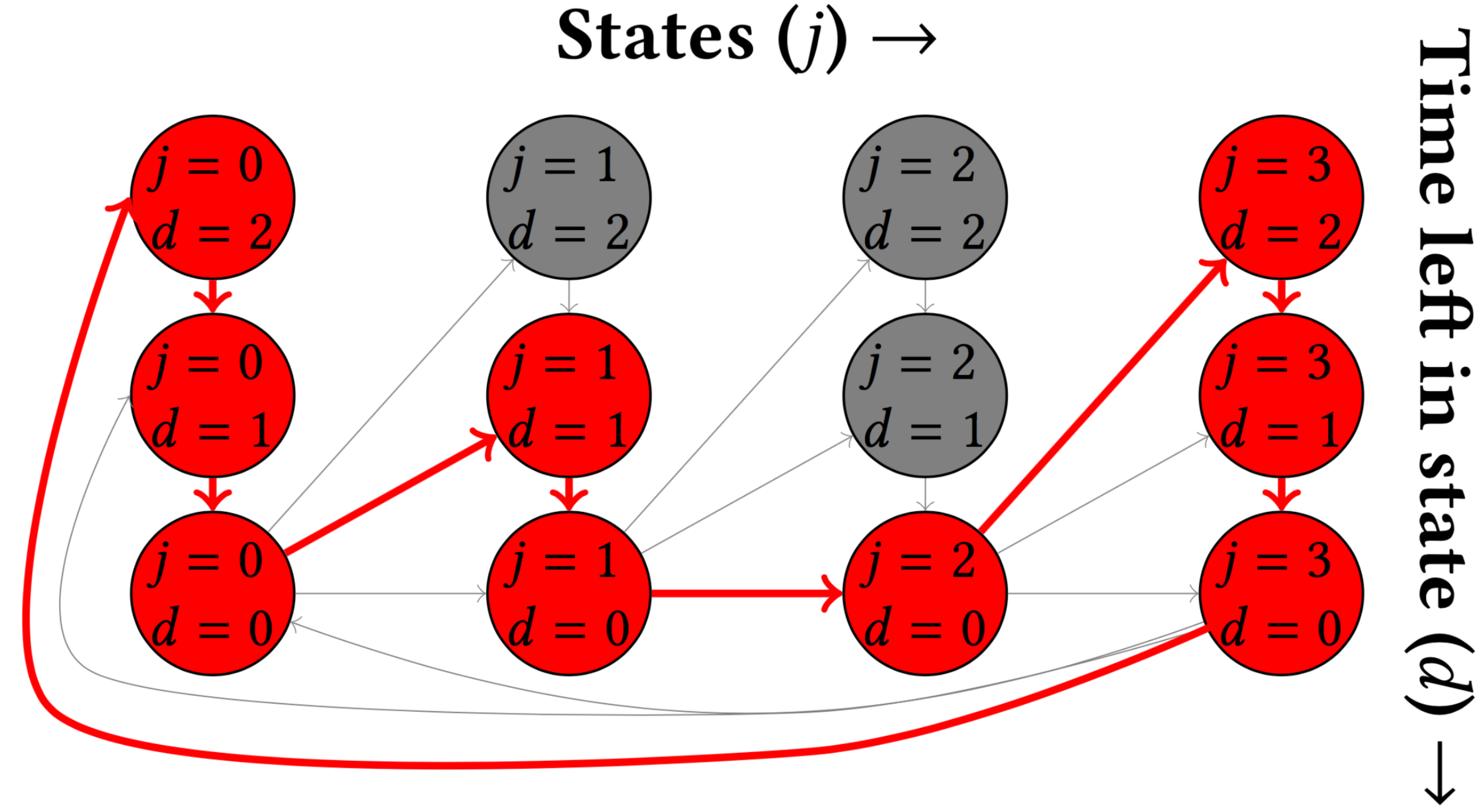}
\caption{Latent states in a CyHMM with J = 4 states and maximum state duration $d_{max} = 2$. The red circles and arrows illustrate one possible path through the CyHMM over the course of a single cycle. The $j$-th latent state counts down until the remaining duration in the state, $d$, is zero and then transitions to the next latent state. All substates with equal $j$ have the same emission parameters. 
} 
\label{fig:CyHMM_diagram}
\end{figure}

The second difference between our model and an ordinary HMM is that, when an individual enters a new state $Z_j$, they also draw a number of timesteps---a \emph{duration}---to remain in that state before transitioning to the next state, an idea from hidden semi-Markov models \cite{yu2010hidden} (HSMMs) and explicit duration hidden Markov models \cite{levinson1986continuously, ferguson1980variable} (EDHMMs). In ordinary HMMs, the time spent in a state is drawn from a geometric distribution  \cite{dewar2012inference}, but as we discuss below, this parameterization is a poor approximation when modeling many real-world cycles; drawing state durations from arbitrary duration distributions allows for more flexible and realistic cycle modeling. 
Figure \ref{fig:CyHMM_diagram} illustrates a path an individual might take through hidden states over the course of a single cycle. Each state $Z_j$  is split into substates indexed by $d$, where $d$ indicates the amount of time remaining in that state.
Substate $Z_{jd}$ progresses deterministically to $Z_{j(d - 1)}$ unless $d = 0$, in which case it can progress to any substate of state $j + 1$ with probabilities drawn from $\mathcal{D}_{j + 1}$, where $\mathcal{D}$ is the duration distribution. (If $j = J$, the model progresses instead to $Z_1$, completing the cycle). All substates of state $Z_j$ share the same emission distribution $\mathcal{E}_j$, and all transitions are either deterministic or parameterized by the duration distributions $\mathcal{D}_j$, so the model's emissions and transitions are completely parameterized by $\mathcal{E}_j$ and $\mathcal{D}_j$\footnote{While in theory some duration distributions can allow an infinite number of substates for each state, in practice duration distributions have negligible mass beyond a certain maximum duration $d_{max}$, so the number of substates can be safely truncated.}. Expressing the model in this way is a commonly used computational technique~\cite{yu2010hidden}, facilitating fast parameter fitting through optimized HMM packages \cite{schreiber2017pomegranate}.

\subsection{Model specification} 

Specifying a CyHMM requires specifying the emission distributions $\mathcal{E}$, the duration distributions $\mathcal{D}$, and the number of states $J$. We now describe and motivate our distributional choices. We find in our experiments that the distributions described below are sufficiently flexible to capture both simulated and real-world data, although our framework extends to other distributions.

\xhdr{Emission distributions} Human activity time series have frequent missing measurements caused, for example, by a user forgetting to track some event on a given day. We show in our real-world case studies (Section \ref{case_studies}), that the probability of missing data fluctuates over the course of the cycle and is important to model. Consequently, we use emission distributions which allow for missing data. We model the data using a mixture where each observation has some probability of being missing and is otherwise drawn from a Gaussian distribution (for continuous features) or a Bernoulli distribution (for binary features)\footnote{Categorical features could be modeled using a simple extension of our Bernoulli model; as there are no categorical features in our real-world datasets, we do not consider them here.}. 

\noindent
\emph{Continuous features}: if a user is in the $j$-th latent state at timestep $t$, we assume their value for the $k$-th feature, $X_{tk}$, is drawn as follows:

 $$   X_{tk} = \left\{
    \begin{array}{ll}
        \rm{unobserved}  & \mbox{if } \text{Bernoulli}(p^\text{obs}_{jk}) = 0 \\
       v_{tk} \sim \mathcal{N}(\mu_{jk}, \sigma_{jk}^2) & \mbox{otherwise}
    \end{array} 
\right .
$$

Hence, $X_{tk}$ is unobserved with some probability $1 -  p^\text{obs}_{jk}$ and otherwise drawn from a Gaussian whose parameters are specific to that state and feature. The parameters for the $j$-th latent state and $k$-th feature are $\mu_{jk}, \sigma_{jk}$, and $p^\text{obs}_{jk}$. Each feature is independent conditional on the hidden state: the emission probability for all features is the product of the emission probabilities for each feature. 

\noindent
\emph{Binary features}: Binary features in real-world time series introduce a subtlety: if all features are zero, it is unclear whether they are true zeros or missing data. For example, if a person on a negative mood-tracking app records no negative mood symptoms, it is unclear whether they were content or whether they merely neglected to log their negative mood. (With continuous features, it is generally more obvious when data is missing, since zero values are out of range). Thus, we assume the following data-generating process: with some probability at each timestep, the person does not bother to log any features and data for all features is missing. Otherwise, they log at least one feature, and we assume they did not experience the features they did not log. Hence, if a person is in the $j$-th latent state at the $t$-th timestep, observed data is drawn as follows: 
$$I^{\text{logged anything}}_t \sim \text{Bernoulli}(p^\text{obs}_j)$$
 $$   X_{tk} = \left\{
    \begin{array}{ll}
        \rm{unobserved}  & \mbox{if } I^{\text{logged anything}}_t = 0 \\
         v_{tk} \sim \text{Bernoulli}(E_{jk}) & \mbox{otherwise}
    \end{array} 
\right .
$$
The parameters of the emission distribution are $p^\text{obs}_j$ and $E_{jk}$. 

\xhdr{Duration distributions}  For ordinary HMMs, the time spent in each state follows a geometric distribution. However, this parameterization does not describe many real-world cycles. 
For example, the length of the luteal phase in the human menstrual cycle is better described by a normal distribution \cite{lenton1984normal}.  In particular, the monotonicity of the geometric probability mass function makes it a poor approximation to many real-world distributions. Instead, we use Poisson distributions for the duration distributions $\mathcal{D}_j$, so the duration distribution for the $j$-th latent state is parameterized by a single rate parameter $\lambda_j$. We use Poisson distributions because they are frequently used in HSMMs, they are fast to fit and easy to interpret, and we show empirically that they provide a better fit to real-world cycles than does an implementation with a geometric distribution. Our framework extends to other distributions. 

\xhdr{Choosing the number of hidden states} $J$ can be chosen either based on prior knowledge about the cycle being modeled (as we do in our case studies) or using cross-validated log likelihood \cite{celeux2008selecting}.

\subsection{Model fitting} 

The optimization procedure is an expectation-maximization algorithm often used with HMMs; we outline it briefly.
\begin{itemize}
\item \textbf{E-step}: Use the Baum-Welch algorithm \cite{bilmes1998gentle} to infer the probability of being in a given latent state at a given timestep, $p_j^{(t)}$, and the expected number of transitions between each pair of substates $Z_{jd}$ and $Z_{j'd'}$. Because of the model structure (Figure~\ref{fig:CyHMM_diagram}), the expected number of transitions is trivial for all start substates with $d \neq 0$ (since the substate must simply count down to the next substate). Denote by $C^{(i)}_{jd}$ the expected number of transitions in time series $i$ into substate $Z_{jd}$ from the prior substate $Z_{(j - 1)0}$.
\item \textbf{M-step}: Update the parameters of the duration distribution $\mathcal{D}_j$ and the emission distribution  $\mathcal{E}_j$  for each latent state.
\begin{itemize} 
\item Updating $\mathcal{D}_j$: In the case of the Poisson distribution, each state has a single rate parameter $\lambda_j$ which is estimated as the mean expected duration in that state: 
$$\lambda_j  = \frac{\sum_{i = 1}^N\sum_{d = 1}^{D} C^{(i)}_{jd} \cdot d}{\sum_{i = 1}^N\sum_{d = 1}^{D} C^{(i)}_{jd}}$$
We set the  probability of beginning in each substate $Z_{jd}$ to be $f_j(d)$, where $f_j$ is the duration probability mass function for the $j$-th latent state.
\item Updating $\mathcal{E}_j$. We update the emission distribution parameters by computing their sufficient statistics. For discrete distributions, the sufficient statistics are the sum of weights  $p_j^{(t)}$ over non-missing samples, the sum of weights over all samples, and the weighted sum of each feature over non-missing samples. For continuous distributions, we must also compute the weighted sum of the square of each feature over non-missing samples. The updates are straightforward; we omit them due to space constraints and provide our implementation online.
\end{itemize}
\end{itemize}

\xhdr{Violation of model assumptions} Model fitting will converge regardless of whether there are actually cycles in the data, although convergence may be slow if model assumptions are seriously violated. Fit should be assessed by comparing model parameters and the outputs discussed in Section \ref{section:cycle_parameter_inference} with prior knowledge about the cycle to determine the plausibility of the fitted model. 

\xhdr{Parameter initialization} CyHMMs are initialized with a hypothesized cycle length. In our simulations (Section \ref{simulations}), we found that CyHMMs are fairly robust to inaccurate specification of this parameter: they converged to the true length as long as they were initialized within roughly 50\% of the true length. In real-world datasets, this is a reasonable constraint: the true cycle length is often at least approximately known (as it is in the two real-world case studies we consider in Section \ref{case_studies}). 
To further increase robustness, the model can be fit using a range of cycle length initializations, and the model with the best log likelihood can then be used; we confirm on simulated data that this works reliably.

\begin{figure}[h!]
\centering
\includegraphics[width=.8\columnwidth]{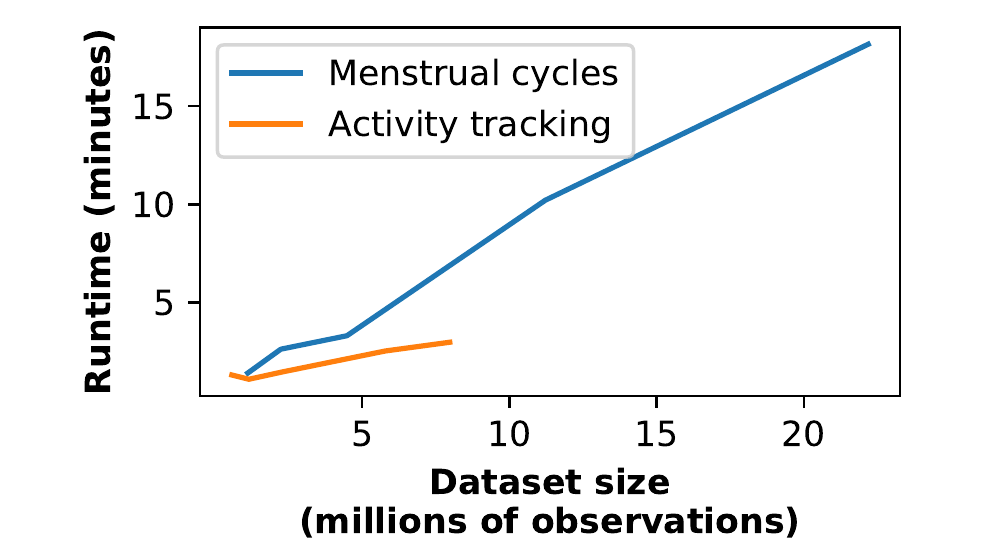}
\caption{Time to fit CyHMM models on our real-world datasets. To provide runtimes on a range of sample sizes, we take random subsamples of the dataset. The menstrual cycles dataset is larger, so we can extrapolate further.} 
\label{fig:CyHMM_timings}
\end{figure}

\xhdr{Scalability} Three features allow our implementation to scale to real-world datasets: we specifically adapt our implementation to use an optimized HMM library \cite{schreiber2017pomegranate}; model fitting can be easily parallelized because the forward-backward computations take up most of the computation time and are independent for each time series; and we find that in real-world applications a relatively small number of hidden states suffices to capture realistic dynamics. Consequently, as we describe in Section \ref{sec:case_studies}, we are able to fit a CyHMM model in 3 minutes (using a computer with 16 cores) on an activity dataset 8 million total observations; we are able to fit a menstrual cycles dataset with 22 million total observations in 18 minutes (Figure \ref{fig:CyHMM_timings}). (By ``observation'', we here mean a reading of one feature for one individual at one timestep). Both these datasets were provided by companies with large user bases and we needed to perform no filtering for scalability; all filtering was performed only to select active users who could provide rich data. For extremely large datasets, a random sample of users could also be used to fit the model, since statistical power would not be a concern. 

\subsection{Inference of cycle properties}
\label{section:cycle_parameter_inference}

\xhdr{Cycle length} To infer cycle length for each individual, we fit a CyHMM to the entire population, then use the fitted model to infer the Viterbi path---that is, the most likely sequence of hidden states---for each individual from their observed data. We define the inferred cycle length as the time to return to the same hidden state in the Viterbi path. 

\xhdr{Feature trajectories} We use CyHMMs to infer how features fluctuate (increase or decrease) over the course of a cycle as follows. After fitting the CyHMM, we assume all users are in a single substate and propagate the state vector forward by multiplying by the learned transition matrix. We compute the expected value of each feature at each timestep using the learned emission model. 

\xhdr{Feature variabilities} We define feature variability to be how much a feature fluctuates over the course of a cycle and compute it as follows. Given the inferred trajectory for feature $k$---a value $V_{tk}$ at each timestep $t$ less than or equal to the cycle length $L$---we define the feature variability $\Delta_{k}$ as how much the feature varies relative to its mean $\mu_k$: $\Delta_{k} = \frac{1}{L}\sum_{t = 1}^{\text{L}} \big|\frac{V_{tk} - \mu_k}{\mu_k}\big| $. (We divide by the mean to ensure features with different scales can be compared; this  definition of $\Delta_k$ is similar to the coefficient of variation \cite{brown1998coefficient}.) Our definition of feature variability is not equivalent to merely computing the standard deviation of all observations of a feature: for example, if our feature is ``person carries umbrella'', its variability over the menstrual cycle will be zero because rain (and umbrellas) do not vary with hormone state. 

\xhdr{Clustering} Previous work has found that clustering time series can identify important heterogeneity \cite{li2011time}; we therefore extend our model to divide individuals into clusters with similar cycle progression patterns. After initializing the clustering by dividing individuals into $C$ clusters\footnote{To initialize the clustering, we choose a random subset of $\mathcal{S}$ individuals and fit one CyHMM $\mathcal{M}_s$ for each individual. We then loop over all individuals in the dataset and compute the log likelihood $\mathcal{L}_{s}^{(i)}$ for each individual $i$ under each model $\mathcal{M}_s$. We run k-means on the matrix of these likelihoods to initially divide individuals into $C$ clusters (z-scoring each individual to control for the fact that, for example, individual time series vary in length and thus log likelihoods may not be directly comparable). This initialization procedure is similar to \cite{smyth1997clustering} but their procedure is quadratic in the number of individuals, rendering it too slow for large real-world datasets (Section \ref{case_studies}).}, we use an EM procedure with hard cluster assignment, alternating between two steps until convergence: 1) for each cluster $c$, fit a separate CyHMM $\mathcal{M}_c$ using only individuals in that cluster; 2) reassign each individual $i$ to the cluster $c$ whose model maximizes the log likelihood of $X^{(i)}$.

\section{Evaluation on Simulated Data}
\label{sec:synthetic_data}

\label{simulations}
We first validate the performance of CyHMMs on synthetic data before examining their application to two real-world datasets.

\subsection{Synthetic data generation}  

We simulate a population where the data for each individual consists of a multivariate time series with missing data. We briefly describe the simulation procedure here and make the full implementation available online. To test our model's robustness to model misspecification, we do not generate data using the CyHMM generative model. In our simulation, each individual progresses through a cycle, and both the values for each feature and the probability that data for the feature is missing vary \emph{sinusoidally} as a function of where the individual is in their cycle. We examine the effects of altering the maximum individual time series length ($T_{\text{max}}$), the amount of noise in the data ($\sigma_n$), the probability data is missing ($p_{\text{missing}}$), and the between-user and within-user variation in cycle length ($\sigma_b$ and $\sigma_w$). We generate simulated data in two stages. First, we simulate the fraction of the way each person is through their cycle at each timestep, which we refer to as ``cycle position";  cycle position can be thought of as a continuously varying cycle state. Second, we generate observed data given the person's cycle position.

\xhdr{Simulating cycle position} We draw each individual's average cycle length from $L^{(i)} \sim N(L, \sigma_b^2)$, where $L$ is the average population cycle length and $\sigma_b$ controls \emph{between individual} variation in cycle length; we set $L = 30$ days in our simulations. We draw each individual's starting cycle day from a uniform distribution over possible cycle days; each subsequent day, their cycle day advances by 1, unless they have reached their current cycle length, in which case they go back to 0 and we draw the length for their next cycle from $N(L^{(i)}, \sigma_w^2)$. $\sigma_w$ controls \emph{within individual} variation in cycle length. We define an individual's \emph{cycle position} $\phi^{(i)}_t \in [0, 1]$ to be the fraction of the way the individual is through their current cycle: \ie, cycle day divided by current cycle length.

\xhdr{Simulating time series given cycle position} Given individual $i$'s cycle position at timepoint $t$, $\phi^{(i)}_t$, we generate the values for a feature $k$, $X_{tk}^{(i)}$, as follows. We generate both binary and continuous data to evaluate our model under a wide variety of conditions. For continuous features, we allow the probability each feature is observed at each timestep, $p_{tk}^{(i)}$, to vary sinusoidally in $\phi^{(i)}_t$. 
We draw whether the feature is observed from $\text{Bernoulli}(p_{tk}^{(i)}$). If the feature is observed, we set its value to a second sinusoidal function of $\phi^{(i)}_t$.
For both sinusoids, we draw individual-specific coefficients to allow for variation in cycle dynamics across individuals. Our emission model for binary features is similar to the continuous model except that at each timestep, a single Bernoulli draw determines whether \emph{all} features are missing, and we draw each feature from a Bernoulli whose emission probability varies sinusoidally in cycle position. Our full procedure for generating simulated data is available online.

\subsection{Inference of cycle length} 

A basic question about any cycle is its length, both in individuals and in the population as a whole. We thus first evaluate how well CyHMMs infer the true cycle length for each individual as compared to baselines. Our evaluation metric is mean error.

\label{baselines}
\xhdr{Baselines} Based on the previous literature we describe in Section \ref{sec:related}, we compare to baselines from all three major areas of related work: classical techniques (Fourier transform and autocorrelation), data-mining techniques (Partial Periodicity Detection) and circadian rhythm detection (the ARSER algorithm): 

\begin{itemize}
\item \textbf{Discrete Fourier transform} \cite{bracewell1986fourier}: for each feature, we perform a discrete Fourier transform  and take the period of the peak with the largest amplitude. We take the median of these periods across all features. (Fourier transforms are generally used on univariate data; by taking the median, we combine data from all features.) 
\item \textbf{Autocorrelation} \cite{pindyck1998econometric}: for each feature, we compute the lag which produces the maximum autocorrelation in that feature. We take the median of these lags across all features. 
\item \textbf{Partial Periodicity Detection} \cite{ma2001mining} (binary data only): we apply Ma and Hellerstein's $\chi^2$ test for partial periodicity detection\footnote{https://github.com/jcborges/PeriodicEventMining} to each individual feature. The algorithm returns a list of statistically significant periods; we take the median of these periods. The algorithm relies on specification of a $\delta$ parameter, which controls the time tolerance in period length; we evaluate $\delta = 2, 5, 10$, find the algorithm yields the most accurate cycle length inference with $\delta = 2$, and report results for this parameter setting. 
\item \textbf{ARSER circadian rhythm detection} \cite{yang2010analyzing}: the ARSER algorithm fits a sinusoidal model to each individual feature\footnote{https://github.com/cauyrd/ARSER}, and reports a $p$-value for periodicity. Consistent with the original authors, we filter for features which show statistically significant periodicities after multiple hypothesis correction; we then take the period with the largest amplitude for each feature and aggregate across multiple features by taking the median. We find that setting a statistical significance threshold of $p = .01$ yields the most accurate cycle length inference, and report results for this parameter setting. We note that the original implementation does not scale to datasets of our size, so to evaluate it we must make two modifications: we parallelize it to use multiple cores, and we fit the autoregression models using only the Yule-Walker and Burg methods, because the MLE method does not scale. 

\end{itemize}

In assessment, we filter out periods which are shorter than 5 days or longer than 50 days (the true cycle length is 30 days). This increases accuracy by, for example, preventing the autocorrelation baseline from returning implausibly short lags (which in real-world settings an analyst would filter out) which degrade performance. We run 100 binary simulation trials and 100 continuous simulation trials, systematically varying the maximum individual time series length ($T_{\text{max}}$) from 90 to 180 days, the amount of noise in the data ($\sigma_n$) from 5 to 50, the probability data is missing ($p_{\text{missing}}$) from 0\% to 90\%, and the between-user and within-user variation in cycle length ($\sigma_b$ and $\sigma_w$) from 1 to 10 days.

\xhdr{Results} CyHMMs have lower mean error than do baselines on both binary and continuous data (Table \ref{tab:simulation_results}). Across all 200 trials, they reduce the mean error in cycle length inference over the baseline with lowest mean error, autocorrelation, by 58\%. Their inferred cycle lengths for each person are also better correlated with the true cycle lengths for each person ($r = .57$) than the baseline with the best correlation (ARSER, $r = .19$) demonstrating that they can better identify individual heterogeneity. 

\begin{table}[h!]
  \resizebox{1\columnwidth}{!}{%
\begin{tabular}{p{4cm}p{1.5cm}p{1.3cm}p{1.5cm}}
\toprule
{} &  All trials &  Binary &  Continuous \\
\midrule
\textbf{CyHMM}                               &          \textbf{2.5 days}  &        \textbf{3.1 days}  &            \textbf{1.9 days}  \\
Autocorrelation                     &          5.9 (58\%) &        6.8 (55\%) &            4.9 (62\%) \\
ARSER                               &         10.5 (76\%) &       13.7 (78\%) &            7.2 (74\%) \\
Fourier                             &         13.3 (81\%) &       16.0 (81\%) &           10.6 (83\%) \\
Partial Periodicity Detection       &         Binary only &       19.4 (84\%) &           Binary only \\
\bottomrule
\end{tabular}
}
\caption{Mean error in inferring cycle length across simulation trials. CyHMMs have lower mean error than all baselines. Relative error improvement of CyHMM over baselines shown in parentheses. We order methods by their mean error. Partial Periodicity Detection can only be applied to binary data.}
\label{tab:simulation_results}
\end{table}

When we examine the effects of altering specific simulation parameters, we find that CyHMM performance, as measured by correlation with true cycle length, improves as noise decreases, fraction of missing data decreases, maximum time series length increases, variability in cycle length \emph{between} individuals increases, and variability in cycle length \emph{within} individuals decreases. We also find that the superior performance of CyHMMs over baselines occurs in part because CyHMMs are more robust to noise and missing data. In very low noise settings, several baselines perform comparably to CyHMMs (autocorrelation, ARSER, and Fourier), but baseline performance rapidly degrades in more realistic scenarios as noise increases. It is logical that CyHMMs would be more robust to noise, because they share information across individuals and across features, while baselines do not. Similarly, CyHMMs should be more robust to missing data because they explicitly incorporate it into the generative model and use it to infer cycle state. We also note that it is striking that CyHMMs can outperform ARSER and Fourier transforms even though both are based on sinusoidal models, and our simulated data is drawn from sinusoids. 

\subsection{Inference of feature trajectories}

We compute feature trajectories and variabilities as described in Section \ref{section:cycle_parameter_inference}, and assess how well the variabilities computed from the inferred trajectories correlate with the variabilities computed from the true trajectories. (To compute the true trajectories, we compute the mean value of each symptom on each cycle day.) The mean correlation between the true and inferred variabilities is 0.97 across all binary simulations and 0.98 across all continuous simulations, demonstrating that CyHMMs are able to correctly infer feature variabilities. Because the baselines discussed in Section \ref{baselines} are not generative models designed to infer feature trajectories, we do not compare to them.

\section{Evaluation on Real-World Data}
\label{sec:case_studies}
\label{case_studies}
We demonstrate the applicability of CyHMMs by using them to infer cycle characteristics in two real-world health datasets: a menstrual cycles dataset, and an exercise and activity tracking dataset. We choose these datasets for two reasons: first, they have information on true cycle state, allowing us to compare our model to existing baselines, and second, they track important health-related features\footnote{Data from both companies has been used in previous studies and users are informed that their data may be used. Because all data analyzed is preexisting and de-identified, the analysis is exempt from IRB review.}. 

\subsection{Datasets} 

\xhdr{Menstrual cycles dataset} The human menstrual cycle is fundamental to women's health; features of the cycle have been linked to cancer \cite{garland1998menstrual}, depression \cite{endicott1993menstrual}, and sports injuries \cite{wojtys1998association}, and the cycle has been proposed as a vital sign \cite{american2006menstruation}.
We obtained data from one of the most popular menstrual cycle tracking apps. Upon logging into the app, users can record \emph{binary} features including period/cycle start (\eg, light, moderate, or heavy bleeding), mood (\eg, happy, sad, or sensitive), pain (\eg, cramps or headache) or sexual activity (\eg, protected or unprotected sex). We define the start of each menstrual cycle as the start of period bleeding. We confirmed that statistics of the dataset were consistent with existing literature on the menstrual cycle: the average cycle duration was 28 days, consistent with previous investigations, and features that peaked before the cycle start were consistent with previous investigations of premenstrual syndrome \cite{yonkers2008premenstrual,pearlstein2005pretreatment}. 

This dataset has a number of traits which make it an useful test case for cycle inference methods. It has a rich set of features for each user, a large number of users, and variation in cycles both between and within users: menstrual cycle length varies from person to person, and even from cycle to cycle, as do cycle symptoms. Most importantly, it contains ground truth on the true cycle start times for each user (as indicated by bleeding). While CyHMMs are more generally applicable to cases when the true cycle starts are not known, true cycle starts are essential for validating that the method works and comparing it to other methods.

We fit a CyHMM using features shown by the tracking app's default five categories---mood, sleep, sex, pain, and energy---because many of the other features had very sparse data. Importantly, we did not provide the model with any features directly related to cycle start, like bleeding, so the model was given no information about when cycles started. In total this left us with 19 binary features for each day. We filtered for users that logged a feature using the app at least 20 times and had a timespan of at least 50 days where they used the app regularly (logging at least once every two days). After filtering, our analysis included 22 million observations (\ie, measurements of one feature for one user at one timestep) from 9,885 app users. We fit a CyHMM with four hidden states because the menstrual cycle is often divided into four latent hormonal phases  \cite{berga1990circadian}. (To study the robustness of our model to this parameter, we also examined results using three or five states.) If a user logged no features on a given day, we considered data for that day to be unobserved because, as explained above, we did not know whether the user truly had no symptoms or merely neglected to log them. 

\xhdr{Activity tracking dataset} We obtained data from a large activity tracking mobile app that has been previously used in studies of human health~\cite{althoff2017onlineactions,althoff2017large,shameli2017gamification}. For each user on each day, our dataset included sleep start time, sleep end time, steps taken, and total calories burned. Each feature had missing data caused, for example, by a user forgetting to log their sleep. Our analysis included all regular app users that logged into the app to record a measurement at least 20 times, had a timespan of at least 50 days where they used the app regularly (logging at least once every two days) and had less than 50\% missing data in every feature; in total, after filtering, our data consisted of 8 million observations from 6,882 users.  Because we are interested in cycles, we removed long-term time trends (\eg, people using the app often lose weight) by subtracting off the centered moving average using a two-week window around each day. This is analogous to the trend-cycle decomposition often used in time-series analysis \cite{fellner1956trends}. 

Previous research has found that people's activity patterns show weekly cycles: for example, they sleep and wake later on the weekends \cite{monk2000sleep}. We thus assessed how well CyHMMs could model this weekend effect from time series data without being provided with the day of the week: essentially, testing whether CyHMMs could recover weekly cycle patterns without knowing the day of the week. We fit a Gaussian CyHMM with two states to correspond to weekday and weekend using sleep start time, sleep end time, steps taken, and calories burned as continuous features. We verified that these features did in fact show weekly cycles in the data: individuals go to bed and wake up later on average on the weekends, with steps taken and calories burned remaining fairly constant across weekends and weekdays. 

\subsection{Analysis of menstrual cycles} 
We used ground truth data to evaluate how well CyHMMs could recover basic features of the menstrual cycle: cycle length, trajectory of each feature over the course of the cycle, most variable features, and clusters of individuals with different symptom patterns. 

\begin{table}[h!]
\resizebox{1\columnwidth}{!}{%
\begin{tabular}{p{4cm}p{2.2cm}p{1.2cm}p{1cm}}
\toprule
Method & Population cycle length & Median error & Mean error \\
\midrule
Ground truth & 28.0 days & 0.0 days & 0.0 days   \\
\textbf{CyHMM}       &              \textbf{29.0} &        \textbf{2.0} &      \textbf{3.6} \\
Autocorrelation             &              19.0 &        9.0 &      9.9 \\
Partial Periodicity Detection  &              21.9 &        8.4 &     10.0 \\
Fourier transform                         &              14.8 &       12.7 &     11.9 \\
ARSER                       &              20.9 &       13.3 &     12.7 \\
\bottomrule
\end{tabular}
}
\caption{Comparison of our model to baselines on menstrual cycle data for users who recorded five or more cycles. Errors reported are averaged across all users; the error for each user is the difference between the true cycle length for each user and the inferred cycle length for that user. We order methods by their mean error.
}
\label{tab:menstrual-baselines}
\end{table}

\subsubsection{Inference of cycle length} As with simulated data, we assessed how well CyHMMs could infer the true cycle length for each user on the menstrual cycles dataset, defining the true cycle length for each user to be their average gap between period starts. To ensure we had reliable data for each user, we assessed performance on users who recorded at least 5 cycles. We compared our model to the baselines described in Section \ref{baselines}. CyHMMs inferred the cycle length for each user more accurately than did the baselines (Table~\ref{tab:menstrual-baselines}) and were also closer to the population average cycle length\footnote{We report results using the 4-state CyHMM, but also found that 3 and 5 state CyHMMs significantly outperformed baselines. We also compared to the baseline of a CyHMM with a geometric duration distribution (rather than a Poisson duration distribution). The errors in inferred cycle length using a geometric rather than a Poisson distribution were 10\%, 47\%, and 129\% higher for the 3, 4, and 5 state models, respectively, confirming that the Poisson distribution provided a better fit to the data.}. CyHMMs had a mean error that was 63\% lower, and a median error that was 78\% lower, than autocorrelation, the baseline with the lowest mean error.

\begin{figure}[t!]
\centering
\includegraphics[width=1\columnwidth]{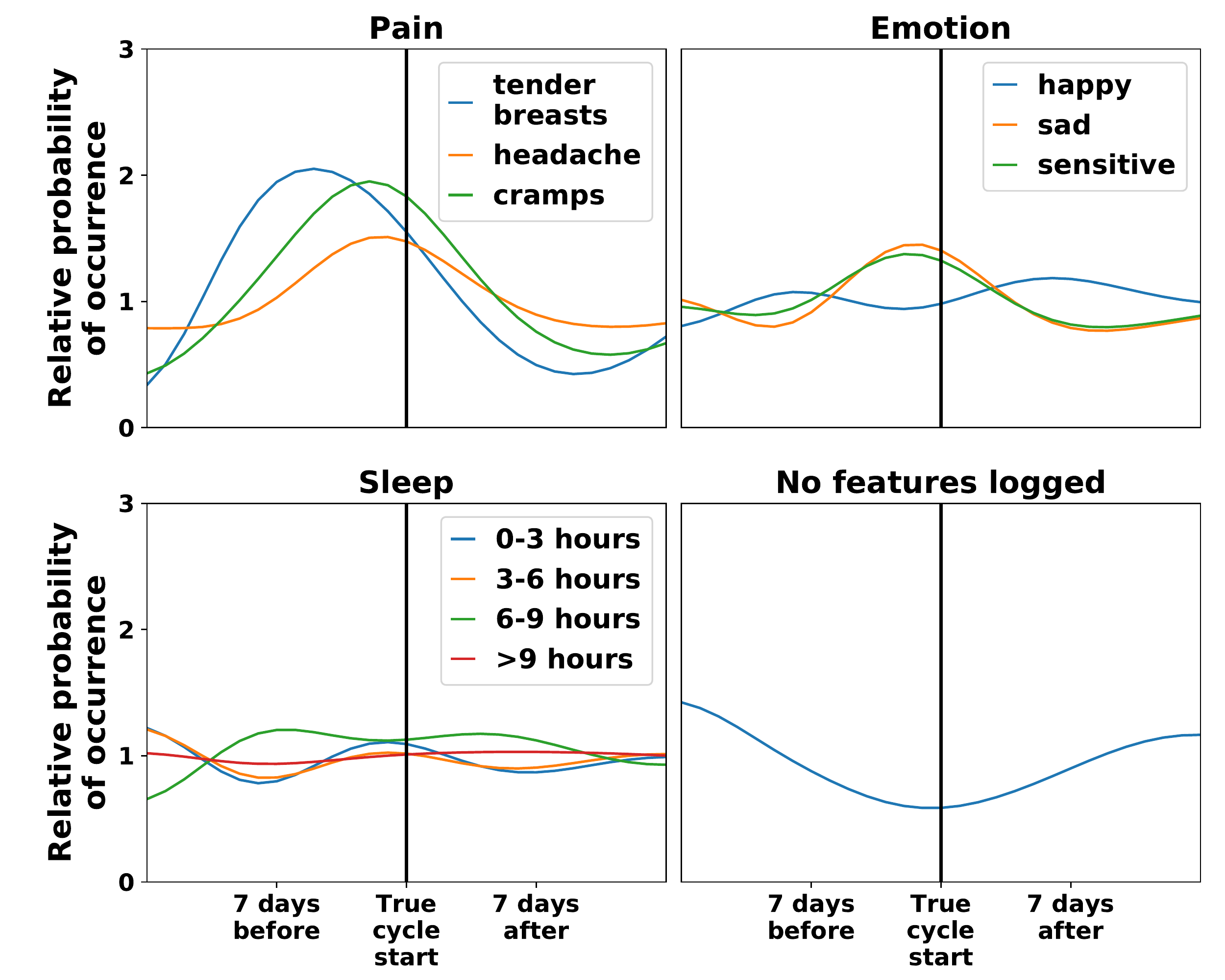}
\caption{Model-inferred feature trajectories for a subset of menstrual cycle features. The vertical axis is the relative probability of emitting a feature (conditional on non-missing data); the horizontal axis is cycle day. For context, the vertical black line shows the day of the true cycle start, although this information is not given to the model. We center cycle start for ease of visualization. Pain and negative mood features increase just prior to cycle start; sleep features do not show consistent patterns. The relative probability that \emph{no} features are logged (``No features logged'') decreases near cycle start, because many people use the app to track their cycle starts. } 
\label{fig:symptom_trajectories}
\end{figure}

\xhdr{Inference of feature trajectories} We inferred feature trajectories using the methodology described in Section ~\ref{section:cycle_parameter_inference}. In Figure ~\ref{fig:symptom_trajectories} we visualize the inferred feature trajectories for pain, emotion, and sleep features. To provide context for the trajectories, we also plot the true cycle start day (when period bleeding starts), although this information is not given to the model. The model correctly infers that mood and pain features increase just before the cycle starts, consistent with existing literature on premenstrual syndrome \cite{yonkers2008premenstrual, pearlstein2005pretreatment}. Finally, the model infers that users are more likely to log in to record features when their cycle is starting; this is logical because many people use the app to track cycle starts, and speaks to the importance of modeling missing data. 

\xhdr{Feature variability} We also found that CyHMMs correctly recovered which features showed the most variability over the course of the cycle. We compare the variabilities computed from the true feature trajectories to the variabilities computed from the inferred feature trajectories as described in Section \ref{section:cycle_parameter_inference}. (We computed the true feature trajectories by aligning all users to their period starts, and computing the fraction of users experiencing a given feature on each day relative to period start). The true variabilities were highly correlated with the inferred variabilities ($r = 0.86$): \eg, the model correctly recovered the fact that pain features showed greater variability than sleep features.

  \begin{table}
  \resizebox{.95\columnwidth}{!}{
\begin{tabular}{llllll}
\toprule
Cluster &     1 &   2 & 3 & 4 & 5 \\
\midrule
Emotion:happy (\% of time)     &   55\% &   15\% &   \textbf{74\%} &   24\% &   66\% \\
Emotion:sad                   &   \textbf{22\%} &    9\% &   11\% &    9\% &   17\% \\
Emotion:sensitive             &   \textbf{46\%} &   15\% &   27\% &   20\% &   32\% \\
High energy                   &   36\% &    7\% &   \textbf{53\%} &   16\% &    4\% \\
Low energy                    &   \textbf{43\%} &   11\% &   33\% &   23\% &    4\% \\
Pain:cramps                   &  \textbf{20\%} &   \textbf{20\%} &   12\% &   12\% &   15\% \\
Pain:headache                 &   \textbf{33\%} &   14\% &   15\% &   14\% &   17\% \\
Pain:tender breasts           &   \textbf{18\%} &   11\% &    9\% &    9\% &   12\% \\
High sex drive                &   \textbf{20\%} &    7\% &    6\% &    6\% &   11\% \\
Had protected sex             &    5\% &   \textbf{15\%} &    4\% &    5\% &    5\% \\
Had unprotected sex           &   10\% &   \textbf{25\%} &    4\% &    8\% &    6\% \\
Had withdrawal sex            &   \textbf{16\%} &    8\% &    2\% &    3\% &    7\% \\
Slept 0-3 hours                  &   \textbf{5\%}  & 1\% &  1\% & 1\% & 0\% \\
Slept 3-6 hours                  &   \textbf{40\%} &    8\% &   18\% &   21\% &    2\% \\
Slept 6-9 hours                   &   39\% &   13\% &   \textbf{63\%} &   52\% &    4\% \\
Slept 9+ hours                   &   10\% &   3\% &   \textbf{13\%} &   9\% &    1\% \\
Features logged per non-missing day            &  \textbf{4.6} &  1.9 &  3.7 &  2.5 &  2.1 \\
Fraction of days with no data                     &    5\% &   \textbf{68\%} &    4\% &   18\% &   10\% \\
\midrule
\midrule
Recorded doctor's appt &   \textbf{37\%} &   15\% &   29\% &   27\% &   15\% \\
Recorded ob/gyn appt   &   \textbf{24\%} &   14\% &   17\% &   18\% &   11\% \\
Took pain medication          &   \textbf{36\%} &   17\% &   27\% &   27\% &   16\% \\
Logged birth control pill            &   38\% &   \textbf{85\%} &   29\% &   49\% &   33\% \\
Negative pregnancy test       &   15\% &   13\% &    9\% &   13\% &   \textbf{16\%} \\
Positive pregnancy test       &    9\% &    5\% &    6\% &    6\% &   \textbf{10\%} \\
Age                           & 21.9 & 21.7 & 20.6 & \textbf{23.9} & 20.8 \\
Cycle length, days            & 28.2 & 29.0 & 28.8 & 28.3 & \textbf{29.4} \\
\bottomrule
\end{tabular}
}
\caption{Clustering of menstrual cycle data. Only the top set of user characteristics (above the double line) is accessible by the model, but the clustering also correlates with differences in medical history, age, and true cycle length (bottom set of characteristics) which are not provided to the model. All differences between clusters are statistically significant ($p < 0.001$, categorical F-test). The largest value in each row is shown in bold.}
\label{tab:menstrual_clustering}

\end{table} 
\xhdr{Clustering of similar individuals} We divided individuals into five clusters with similar feature patterns using the methodology described in Section \ref{section:cycle_parameter_inference}. We chose the number of clusters using the elbow method \cite{tibshirani2001estimating}: beyond five clusters, both the train and test log likelihoods increased much more slowly.  Our clustering correlates with individual-specific features not given to the model (Table~\ref{tab:menstrual_clustering}; all features below the double horizontal line), like medical history and age. We report only features which show statistically significant differences (categorical F-test $p<0.001$), revealing distinct groups: 
 
 \begin{itemize}
 \item \textbf{Cluster 1: Severe symptom users.} This cluster (Table~\ref{tab:menstrual_clustering}, first column), is most likely to report going to the doctor or gynecologist or taking pain medication (rows below double horizontal line). Our model is able to identify this subcluster even though it is given no information on doctors' visits. These users also have the highest rate of pain features, are more likely to report negative emotions and less sleep, and are most likely to report having a high sex drive (rows above double horizontal line). 
  \item \textbf{Cluster 2: Birth control users.} These users primarily use the app to keep track of whether they have taken birth control (``Logged birth control pill'' row)---again, not information given to the model. Hence, they are less likely to log other features and have a high fraction of missing data (``Fraction of days with no data'' row), but likely to be on birth control. Consistent with this, they have a higher rate of having unprotected sex than do other users, and the lowest rate of positive pregnancy tests. 
   \item \textbf{Cluster 3: Happy users.} This group of users is most likely to report positive features like happy emotion (top row), high energy, and sleeping 6-9 hours.
      \item \textbf{Cluster 4: Low-emotion users.} This group of users is relatively unlikely to log any emotion features (top three rows). 
 \item \textbf{Cluster 5: Infrequent loggers.} This cluster of users appears to put less effort into using the app; when they record features, they record the fewest of any group besides the birth control group (``Features logged per non-missing day'' row). 
\end{itemize}

The fact that the clustering correlates with features not used in the clustering---like birth control, doctors' appointments, and age---indicates that CyHMMs are correctly identifying true population heterogeneity. For example, the model is able to identify ``users more likely to be on birth control'' (information it is not given) as ``users with a high probability of missing data'' (something it explicitly models). Recovering the birth control subpopulation is essential for accurately modeling the menstrual cycle. Previous work has found that women on birth control experience different menstrual cycle symptom progression---for example, less variability in mood over the course of the cycle \cite{oinonen2002extent, paige1971effects}. (We note that the fact that women in the birth control cluster are relatively unlikely to bother logging emotion symptoms is consistent with this lack of variability.) Recovering the birth control subpopulation is also important for designing health interventions. For example, women on birth control require different pharmacologic treatment \cite{hassan1987pharmacologic} and, because their hormone cycles are different, have different risks of hormone-related cancers \cite{collaborative1996breast, narod1998oral}.

\subsection{Analysis of cycles in activity tracking}

\begin{figure}[h!]
\centering
\includegraphics[width=.7\columnwidth]{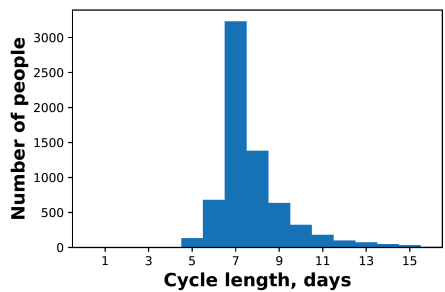}
\caption{Histogram of cycle lengths for individuals in the activity tracking dataset. The cycle length for each individual is the modal cycle length for that individual.}\label{fig:azumio_cycle_lengths}
\end{figure}

\xhdr{Inference of population cycle length} 
We define the model-inferred cycle lengths as in Section \ref{section:cycle_parameter_inference}. Since there are no ground truth cycle lengths for all individuals in this dataset---unlike with the menstrual cycles dataset, there is no clear physiological feature with which to define cycle starts---a supervised comparison to baselines is impossible in this setting. Instead, we directly evaluate the plausibility of CyHMM cycle length inference. When we apply CyHMMs to our dataset, the model infers that the most frequent cycle length in the population as a whole is 7 days (Figure~\ref{fig:azumio_cycle_lengths}). Inspecting the model, the two learned hidden states also clearly correspond to weekday and weekend, even though day of the week is not provided to the model: the ``weekend hidden state'' has an inferred average duration of 2.2 days (as compared to the true weekend length of 2 days), while the ``weekday hidden state'' has an inferred average duration of 4.9 days (as compared to the true weekday length of 5 days).  The weekend hidden state has a later inferred mean for sleep start and sleep end than the weekday hidden state, consistent with the fact that individuals sleep later on the weekends. On timesteps when the model infers an individual is in the weekend state, it is twice as likely to be the true weekend. All these facts indicate that the model is successfully recovering weekday/weekend cycles. This is consistent with previous findings that human activity shows weekly cycles \cite{monk2000sleep}. 

\xhdr{Clustering of similar individuals}
We next investigated whether CyHMMs could cluster individuals into groups that correlated with fundamental features like BMI, age, and gender without being provided with these features to use in the clustering. Our clustering shows statistically significant correlations with all these features ($p < 0.001$, categorical F-test). For example, we identify one cluster with the highest proportion of males (78\%) with the highest BMIs (26.9); another cluster, essentially its opposite, has the lowest proportion of males (48\%) and the lowest BMIs (25.4). (Cluster differences between BMIs remain significant when adjusting for age and sex via linear regression.) None of these features are given to the CyHMM in performing the clustering, indicating that the model is successfully identifying true population heterogeneity. Our finding is consistent with previous work that finds that weekend-weekday sleep differences correlate with important health metrics like body mass index (BMI) \cite{roenneberg2012social}.  CyHMMs can thus identify subpopulations that differ along fundamental health metrics like BMI, potentially allowing for unsupervised discovery of behavioral risk factors to inform monitoring of obesity.

\section{Conclusion}
\label{sec:conclusion}

Modeling cycles in time series data is important for understanding human health, but it is challenging because cycle starts are rarely labeled. Here we present CyHMMs, which take as input a multivariate time series for each individual in a population, infer what latent cycle state an individual is in at each timestep, and use this inferred state to recover fundamental cycle characteristics. Evaluating our method on both simulated data and two real-world datasets with ground-truth information on cycle states, we find CyHMMs infer the true cycle length more accurately than do baselines. CyHMMs can also infer cycle characteristics that baselines cannot: we show they can infer how each observed feature changes over the course of the cycle, accurately recover the most variable features, and find clusters of individuals with distinct cycle patterns. While we evaluated CyHMMs on datasets in which cycle starts are known in order to validate our method, they are designed to be applied to datasets on which cycle starts are unknown: for example, sentiment in individual Twitter feeds.

CyHMMs assume that observed data is generated by a single latent state that progresses cyclically. Future work could attempt to relax this assumption while retaining the interpretability it provides. Natural extensions to the CyHMM model might retain its core idea of a cyclic latent state while increasing the model's expressive power: for example, by using a multidimensional hidden state as factorial HMMs \cite{ghahramani1996factorial} or LSTMs \cite{hochreiter1997long} do, with some cyclic and some unconstrained dimensions. A multidimensional hidden state could capture data that includes multiple cycles, or both a cycle and a time trend. These extensions flow from the fact that CyHMMs, by using a generative model with hidden state, provide a natural and flexible way to model cycles fundamental to human health.

\xhdr{Code availability} https://github.com/epierson9/cyclic\_HMMs. 

\xhdr{Acknowledgments} We thank David Hallac, Chris Olah, Nat Roth, Marinka Zitnik, the apps that provided data, and the reviewers for their valuable feedback. E.P. was supported by Hertz and NDSEG Fellowships. T.A. was supported by National Institutes of Health (NIH) grant U54 EB020405 and a SAP Stanford Graduate Fellowship. This research was supported in part by NIH BD2K, DARPA NGS2, ARO MURI, IARPA HFC, Stanford Data Science Initiative, and Chan Zuckerberg Biohub.

\balance

\bibliographystyle{abbrv}
{\tiny
\bibliography{refs}
}

\end{document}